\documentclass{ws-ijgmmp}
\usepackage{amsmath}

\begin{document}

\title{SYMMETRY ANALYSIS OF THE KLEIN-GORDON EQUATION IN BIANCHI I SPACETIMES%
}
\author{A. PALIATHANASIS}
\address{Dipartimento di Fisica, Universita' di Napoli, ``Federico II'', Compl.
Univ. di Monte S. Angelo, Via Cinthia, I-80126, Napoli, Italy\\
INFN Sez. di Napoli, Compl. Univ. di Monte S. Angelo, Via
Cinthia, I-80126, Napoli, Italy\\
\email{paliathanasis@na.infn.it}}

\author{M. TSAMPARLIS}
\address{Faculty of Physics, Department of Astrophysics - Astronomy -
Mechanics,University of Athens, Panepistemiopolis, Athens 157 83,
Greece\\
\email{mtsampa@phys.uoa.gr}}

\author{M. T. MUSTAFA}
\address{Department of Mathematics, Statistics and Physics,
College of Arts and Sciences, Qatar University, Doha 2713, Qatar\\
\email{tahir.mustafa@qu.edu.qa}}
\maketitle

\begin{abstract}
In this work we perform the symmetry classification of the Klein Gordon
equation in Bianchi I spacetime. We apply a geometric method which relates
the Lie symmetries of the Klein Gordon equation with the conformal algebra
of the underlying geometry. Furthermore, we prove that the Lie symmetries
which follow from the conformal algebra are also and Noether symmetries for
the Klein Gordon equation. We use these resutls in order to determine all
the potentials in which the Klein Gordon admits Lie and Noether symmetries.
Due to the large number of cases and for easy reference the results are
presented in the form of tables. For some of the potentials we use the Lie
admitted symmetries to determine the corresponding invariant solution of the
Klein Gordon equation. Finally, we show that the results also solve the
problem of classification of Lie/Noether point symmetries of the wave
equation in Bianchi I spacetime and can be used for the determination of
invariant solutions of the wave equation.
\end{abstract}

\markboth{A. Paliathanasis, M. Tsamparlis and M.T. Mustafa}
{Symmetry analysis of the Klein Gordon equation in Bianchi I spacetimes}

%
%

\begin{history}
\received{(Day Month Year)}
\revised{(Day Month Year)}
\end{history}

\section{Introduction}

The study of differential equations in a general Riemannian space involves
investigation of equations containing arbitrary parameters or arbitrary
functions. This naturally leads to the group classification problem of
differential equations. The first group classification problem was carried
out by Ovsiannikov \cite{ovsiannikov} who classified all forms of the
non-linear heat equation $u_{t}=(f(u)u_{x})_{x}$. Studies related to group
properties of non-linear wave equations began with the well-known paper of
Ames \cite{Ames81} in 1981. Since then the group classification problem has
been studied intensely for fundamental equations arising from models in
engineering and physics, e.g. the symmetry classification of the geodesic
equations of Riemannian spaces \cite{TsamNLD,Camci2014}, the symmetry
classification of the two and three dimensional Newtonian systems \cite%
{Damianou2d,Damianou3d,Tsam2d,Tsam3d} and many others \cite{Ibragimov1991,
Torrisi1998, Zhdanov1999, Gandarias2004, Ivanova2004, Lahno2006,
Ivanova2006, Ivanova2007, Mahomed2007, Vaneeva2008, Ivanova2010, Azad2010,
Tracina2010, TsamQadir,Chris,Jamal1,Camci}. Furthermore, a symmetry analysis
of wave equation in a power-law Bianchi III spacetime spacetime can be found
in \cite{Jamal2} and a symmetry analysis of the wave equation on static
spherically symmetric spacetimes, with higher symmetries, was recently
carried out in \cite{Azad2013}

In \cite{AnJGP}, it was proved that for a linear, in the derivatives, second
order partial differential equation (PDE) the Lie point symmetries are
related with the conformal algebra of the geometry defined by the PDE.

One important equation which belongs to this type of PDEs is the Klein
Gordon equation
\begin{equation}
\Delta u+V\left( x^{i}\right) u=0  \label{KG.Eq1}
\end{equation}%
where $\Delta =\frac{1}{\sqrt{\left\vert g\right\vert }}\frac{\partial }{%
\partial x^{i}}\left( \sqrt{\left\vert g\right\vert }\frac{\partial }{%
\partial x^{j}}\right) $ is the Laplace operator defined in terms of the
metric of the Riemannian space.

In \cite{AnIJGMMP} the generic Lie symmetry vector for the Klein Gordon
equation in a Riemannian space has been found and it has been expressed in
terms of the elements of the conformal algebra of the space modulo a
constraint relation involving the Lie symmetry vector and the potential
entering the Klein Gordon equation.

Concerning the Noether point symmetries of the Klein Gordon equation in a
Riemannian space we prove that these are the same as the Lie point
symmetries of this equation.

These geometric results transfer the problem of the Lie / Noether symmetry
classification of the Klein Gordon equation (\ref{KG.Eq1}) in a Riemannian
space to the problem of determining the CKVs of the space and the
appropriate potentials which solve the corresponding constraint condition.
We note that taking the potential $V\left( x^{i}\right) =0$ this
classification provides, as a special case, the corresponding symmetry
analysis for the wave equation in Bianchi I spacetimes.

These results can be useful in two ways. Using the reduction of the Klein
Gordon equation by means of certain Lie point symmetries one determines
invariant solutions of (\ref{KG.Eq1}) with respect to these symmetries.
Furthermore because the Lie symmetries are also Noether symmetries one is
able to determine conserved currents for the corresponding potential.

We apply these general results in Bianchi I spacetime and perform the
complete group classification of the Lie /Noether point symmetries of Klein
Gordon equation in this spacetime in terms of the potential $V\left(
x^{i}\right) $.

The paper is organized as follows. Section \ref{Premil} provides the
geometrical preliminaries and the theoretical background. In section \ref%
{LieNKG}, we give the explicit form of the generic Lie symmetry vector of
the Klein Gordon equation in a general Riemannian space in terms of the
conformal algebra of the space modulo a constraint condition involving the
potential. In section \ref{KGBianchi} we obtain the symmetry classification
for the Klein Gordon equation in Bianchi I spacetimes and the corresponding
potentials. Furthermore, in section \ref{invariant} the Lie point symmetries
are applied in order to reduce the Klein Gordon equation and determine
invariant solutions. Finally in section \ref{conclusions} we draw our
conclusion and we discuss the symmetry analysis of the wave equation.

\section{Preliminaries}

\label{Premil}

In this section we give the basic definitions and properties of the
collineations of spacetime and of the point symmetries of differential
equations.

\subsection{Collineations of Riemannian spaces}

A collineation in a Riemannian space is a vector field $\mathbf{X}$ which
satisfies an equation of the form%
\[
\mathcal{L}_{X}\mathbf{A}=\mathbf{B}
\]%
where $\mathcal{L}_{X}$ is the Lie derivative with respect to the vector
field $\mathbf{X}$, $\mathbf{A}$ is a geometric object (not necessarily a
tensor) defined in terms of the metric and its derivatives and $\mathbf{B}$
is an arbitrary tensor with the same indices as the geometric object $%
\mathbf{A}.$ The collineations of Riemannian spaces have been classified by
Katzin et.al. \cite{Katzin}. In the following we are interested in the
collineations of the metric tensor i.e. $\mathbf{A}=g_{ij}$.

If there exists a function $\psi \left( x^{k}\right) $ so that $\mathbf{B}%
=2\psi \left( x^{k}\right) g_{ij}$ the vector field $X$ is called a
Conformal Killing vector (CKV) if $\psi \left( x^{k}\right) \ne 0 $, a
special CKV (sp.CKV) if $\psi _{;ab}=0$, a homothetic (HV) if $\psi= $%
~constant and a Killing vector (KV) if $\psi =0$.

The CKVs of the metric $g_{ij}$ form a closed Lie algebra which is called
the conformal algebra of the metric $g_{ij}$. Two metrics $g_{ij},\bar{g}%
_{ij}$ which are conformally related, that is there exist a function $%
N^{2}\left( x^{k}\right) $ such as $\bar{g}_{ij}=N^{2}\left( x^{k}\right)
g_{ij}$, have the same conformal algebra. A space is called conformally flat
if it is conformally related to the flat space,

The conformal algebra contains two closed subalgebras, the Homothetic
algebra and the Killing algebra related as follows%
\[
KVs\subseteq HVs\subseteq CKVs.
\]

The dimension of the conformal algebra of a$~n-$dimensional metric $(n>2)$
of constant curvature is $\frac{1}{2}\left( n+1\right) \left( n+2\right), $
the dimension of the Killing algebra is $\frac{1}{2}n\left( n+1\right) $ and
the dimension of the Homothetic algebra is $\frac{1}{2}n\left( n+1\right)
+1. $ The flat space admits $\frac{1}{2}\left( n+1\right) \left( n+2\right) $
CKVs.

\subsection{Point symmetries of differential equations}

A partial differential equation (PDE) is a function $%
H=H(x^{i},u^{A},u_{,i}^{A},u_{,ij}^{A})$ in the jet space $\bar{B}_{\bar{M}}$%
, where $x^{i}$ are the independent variables and $u^{A}$ are the dependent
variables. The infinitesimal point transformation
\begin{align}
\bar{x}^{i}& =x^{i}+\varepsilon \xi ^{i}(x^{k},u^{B})~,  \label{pr.01} \\
\bar{u}^{A}& =u^{A}+\varepsilon \eta ^{A}(x^{k},u^{B})~,  \label{pr.02}
\end{align}%
has the infinitesimal symmetry generator
\begin{equation}
\mathbf{X}=\xi ^{i}(x^{k},u^{B})\partial _{x^{i}}+\eta
^{A}(x^{k},u^{B})\partial _{u^{A}}~.  \label{pr.03}
\end{equation}

The generator $\mathbf{X}$ of the infinitesimal transformation (\ref{pr.01}%
),(\ref{pr.02}) is called a Lie point symmetry of the PDE $H$ if there
exists a function $\lambda $ such that the following condition holds \cite%
{Ibrag,Stephani}
\begin{equation}
\mathbf{X}^{[n]}(H)=\lambda H~,~modH=0,  \label{pr.04}
\end{equation}%
where
\begin{equation}
\mathbf{X}^{[n]}=\mathbf{X}+\eta _{i}^{A}\partial _{u_{i}^{A}}+\eta
_{ij}^{A}\partial _{u_{ij}^{A}}+...+\eta _{i_{1}i_{2}...i_{n}}^{A}\partial
_{u_{i_{1}i_{2}...i_{n}}^{A}}  \label{pr.05}
\end{equation}%
is the $n^{th}$ prolongation vector and
\begin{equation}
\eta _{i}^{A}=\eta _{,i}^{A}+u_{,i}^{B}\eta _{,B}^{A}-\xi
_{,i}^{j}u_{,j}^{A}-u_{,i}^{A}u_{,j}^{B}\xi _{,B}^{j}~,  \label{pr.06}
\end{equation}%
with
\begin{align}
\eta _{ij}^{A}& =\eta _{,ij}^{A}+2\eta _{,B(i}^{A}u_{,j)}^{B}-\xi
_{,ij}^{k}u_{,k}^{A}+\eta _{,BC}^{A}u_{,i}^{B}u_{,j}^{C}-2\xi
_{,(i|B|}^{k}u_{j)}^{B}u_{,k}^{A}  \nonumber \\
& -\xi _{,BC}^{k}u_{,i}^{B}u_{,j}^{C}u_{,k}^{A}+\eta
_{,B}^{A}u_{,ij}^{B}-2\xi _{,(j}^{k}u_{,i)k}^{A}-\xi _{,B}^{k}\left(
u_{,k}^{A}u_{,ij}^{B}+2u_{(,j}^{B}u_{,i)k}^{A}\right)  \label{pr.07}
\end{align}

Lie point symmetries of differential equations can be used in order to
determine invariant solutions or transform solutions to solutions \cite%
{Bluman}. From condition (\ref{pr.04}) one defines the Lagrange system%
\[
\frac{dx^{i}}{\xi ^{i}}=\frac{du}{\eta }=\frac{du_{i}}{\eta _{\left[ i\right]
}}=...=\frac{du_{ij..i_{n}}}{\eta _{\left[ ij...i_{n}\right] }}
\]%
whose solution provides the characteristic functions
\[
W^{\left[ 0\right] }\left( x^{k},u\right) ,~W^{\left[ 1\right] i}\left(
x^{k},u,u_{i}\right) ,...,W^{\left[ n\right] }\left(
x^{k},u,u_{,i},...,u_{ij...i_{n}}\right) .
\]%
The solution $W^{\left[ k\right] }$ is called the kth order invariant of the
Lie point symmetry vector (\ref{pr.03}). These invariants can be used in
order to reduce the order of the PDE (for details see e.g. \cite{Stephani}).

The Lie point symmetries of a PDE span a Lie algebra $G_{L}$ of dimension $%
\dim G_{L}>1$. \ The application of a Lie symmetry to a PDE $H$ leads to a
new differential equation $\bar{H}$ which is different from $H$ and is
possible to admit Lie symmetries which are not Lie symmetries of $H$ (these
Lie symmetries are called Type II hidden symmeties). It has been shown \cite%
{Govinger} that if $X_{1},X_{2}$ are Lie point symmetries of the original
PDE with commutator $\left[ X_{1},X_{2}\right] =cX_{1}~$where $c$ is a
constant, then reduction by $X_{2}$ results in $X_{1} $ being a point
symmetry of the reduced PDE $\bar{H}$ \ while reduction by $X_{1}$ results
in a PDE $\bar{H}$ which has no relevance for the PDE $\bar{H} $.

For PDEs arising from a variational principle, Noether's theorem states \cite%
{Bluman}.

\begin{theorem}
The action of the generator (\ref{pr.03}) of the infinitesimal
transformation (\ref{pr.01}),(\ref{pr.02}) on the Lagrangian $%
L=L(x^{k},u^{A},u_{k}^{A})$ leaves $H(x^{i},u^{A},u_{,i}^{A},u_{,ij}^{A})$
invariant if there exists a vector field $A^{i}=A^{i}(x^{i},u^{A})$ such
that the following condition is satisfied
\begin{equation}
\mathbf{X}^{[1]}L+LD_{i}\xi ^{i}=D_{i}A^{i}~.  \label{pr.08}
\end{equation}%
The corresponding Noether flow $I^{i}$ is defined by the expression
\begin{equation}
I^{i}=\xi ^{k}\left( u_{k}^{A}\frac{\partial L}{\partial u_{i}^{A}}-\delta
_{k}^{i}L\right) -\eta ^{A}\frac{\partial L}{\partial u_{i}^{A}}+A^{i}~.
\label{pr.09}
\end{equation}%
and it is conserved, that is satisfies the relation~%
\begin{equation}
D_{i}I^{i}=0.  \label{pr.09.1}
\end{equation}
\end{theorem}

In (\ref{pr.09.1}) $D_{i}$ is the total derivative, defined as follows:%
\[
D_{i}=\partial _{x^{i}}+u_{i}^{A}\partial _{u^{A}}+u_{ij}^{A}\partial
_{u_{j}^{A}}+...
\]

\section{Point symmetries of the Klein Gordon equation in a general
Riemannian space}

\label{LieNKG}

In a recent paper \cite{AnIJGMMP} it has been shown that the Lie point
symmetries of the Klein Gordon equation (\ref{KG.Eq1}) in a general
Riemannian space are elements of the conformal algebra of the space. More
specifically the following theorem is proved.

\begin{theorem}
\label{KG}The Lie point symmetries of the Klein Gordon equation (\ref{KG.Eq1}%
)$\ $in a Riemannian space of dimension \ $n$ are generated from the
elements of the conformal algebra of the metric~$g_{ij}$ defining the
Laplace operator, as follows\newline
a) for $n>2$ the Lie symmetry vector is%
\begin{equation}
X=\xi ^{i}\left( x^{k}\right) \partial _{i}+\left( \frac{2-n}{2}\psi \left(
x^{k}\right) u+a_{0}u+b\left( x^{k}\right) \right) \partial _{u}
\label{KGT.01}
\end{equation}%
where $\xi ^{i}$ is a CKV with conformal factor $\psi \left( x^{k}\right) $,$%
~b\left( x^{k}\right) $ is a solution of (\ref{KG.Eq1})~and the following
condition involving the potential is satisfied%
\begin{equation}
\xi ^{k}V_{,k}+2\psi V-\frac{2-n}{2}\Delta \psi =0.  \label{KGT.02}
\end{equation}%
b) for $n=2$ the Lie symmetry vector is
\begin{equation}
X=\xi ^{i}\left( x^{k}\right) \partial _{i}+\left( a_{0}u+b\left(
x^{k}\right) \right) \partial _{u}  \label{KGT.03}
\end{equation}%
where $\xi ^{i}$ is a CKV with conformal factor $\psi \left( x^{k}\right) $,$%
~b\left( x^{k}\right) $ is a solution of (\ref{KG.Eq1}) and the following
condition is satisfied%
\begin{equation}
\xi ^{k}V_{,k}+2\psi V=0.  \label{KGT.04}
\end{equation}
\end{theorem}

The Klein Gordon equation (\ref{KG.Eq1}) follows from the Lagrangian%
\begin{equation}
L\left( x^{i},u,u_{,i}\right) =\frac{1}{2}\sqrt{g}g^{ij}u_{,i}u_{,j}-\frac{1%
}{2}\sqrt{g}V\left( x^{i}\right) u^{2}.  \label{KGL.01}
\end{equation}

For each term of the Noether symmetry condition (\ref{pr.08}) we have%
\begin{equation}
D_{i}A^{i}=A_{,i}^{i}+A_{,u}^{i}u_{,i}  \label{KGL.02}
\end{equation}%
\begin{equation}
LD_{i}\xi ^{i}=L\xi _{,i}^{i}+L\xi _{,u}^{i}u_{,i}  \label{KGL.03}
\end{equation}%
\begin{equation}
\mathbf{X}^{[1]}L=\xi ^{i}\frac{\partial L}{\partial x^{i}}+\eta \frac{%
\partial L}{\partial u}+\eta _{i}\frac{\partial L}{\partial u_{i}}.
\label{KGL.04}
\end{equation}

Therefore by collecting the terms of the same powers of $\left( u_{i}\right)
^{K}$ we find that condition (\ref{pr.08}) is equivalent to the following
determining system of equations%
\begin{equation}
\xi _{,u}^{k}=0~,~\sqrt{g}g^{ij}\eta _{,j}=A_{,u}^{i}  \label{KGL.05}
\end{equation}%
\begin{equation}
-\frac{1}{2}\sqrt{g}\left( \ln \left( \sqrt{g}\right) _{,k}\xi
^{k}Vu^{2}+V_{,k}\xi ^{k}u^{2}-2Vu\eta +\xi _{,k}^{k}Vu^{2}\right)
=A_{,i}^{i}  \label{KGL.06}
\end{equation}%
\begin{equation}
g_{,k}^{ij}\xi ^{k}-2g^{k(i}\xi _{,k}^{j)}+2g^{ij}\eta _{,u}+g^{ij}\xi
_{,k}^{k}+\ln \left( \sqrt{g}\right) _{,k}\xi ^{k}g^{ij}=0.  \label{KGL.07}
\end{equation}

The solution of this system is as follows. Equation (\ref{KGL.05}) implies $%
\xi ^{i}=\xi ^{i}\left( x^{k}\right) $ and $A_{i}=\sqrt{g}\int \eta
_{,i}du+\Phi _{i}\left( x^{k}\right) $. Combining this with $\left( \sqrt{g}%
\right) _{,k}=\sqrt{g}\Gamma _{kr}^{r}$ where $\Gamma _{jk}^{i}$ are the
connection coefficients, condition (\ref{KGL.07}) becomes%
\begin{equation}
\mathcal{L}_{\xi }g_{ij}=\left( \xi _{;k}^{k}+2\eta _{,u}\right) g_{ij}
\label{KGL.08}
\end{equation}%
which means that $\xi ^{i}=\xi ^{i}\left( x^{k}\right) $ is a CKV of the
metric $g_{ij}$ and that $\xi _{;k}^{k}=n\psi \left( x^{k}\right) .$ However
since the Noether point symmetries are also Lie point symmetries, from (\ref%
{KGT.01}) we have that $\eta \left( x^{k},u\right) =\frac{2-n}{2}\psi \left(
x^{k}\right) u+a_{0}u+b\left( x^{k}\right) $; where $\Delta b+V\left(
x^{i}\right) b=0$. Then condition (\ref{KGL.06}) becomes%
\begin{equation}
V_{,k}\xi ^{k}u^{2}+2\psi Vu^{2}+a_{0}Vu=\frac{2-n}{2}\Delta \psi u^{2}-%
\frac{2}{\sqrt{g}}\Phi _{,i}^{i}.  \label{KGL.09}
\end{equation}%
This equation is an identity hence the coefficients of the various powers of
$u$ must vanish. It follows then that $a_{0}=0$, $\Phi _{,i}^{i}=0$ and the
Noether condition becomes
\begin{equation}
V_{,k}\xi ^{k}+2\psi V-\frac{2-n}{2}\Delta \psi =0  \label{KGL.09.1}
\end{equation}%
which is the Lie point symmetry condition (\ref{KGT.02}). We conclude that
every Lie point symmetry (but not the trivial vector field $u\partial _{u}$,
since $a_{0}=0$ ) of the Klein Gordon equation is also a Noether point
symmetry for the Lagrangian (\ref{KGL.01}). Therefore we have the following
theorem

\begin{theorem}
\label{KGNoether}The Noether point symmetries of the Klein Gordon Lagrangian
(\ref{KGL.01}) are generated by the CKVs of the metric $g_{ij}$. The generic
Noether point symmetry vector is the Lie generic point symmetric vector with
$a_{0}=0$ and corresponding gauge function%
\begin{equation}
A_{i}=\frac{2-n}{4}\sqrt{g}\psi _{,i}\left( x^{k}\right) u^{2}+\Phi
_{i}\left( x^{k}\right)
\end{equation}%
where $\Phi _{i}\left( x^{k}\right) $ is a function which satisfies the
condition $D_{i}\Phi ^{i}=0.$
\end{theorem}

From Noether's theorem the function $\Phi _{i}\left( x^{k}\right) $ can be
absorbed in the Noether flow, hence without loss of generality we can set $%
\Phi _{i}\left( x^{k}\right) =0$.

Theorem \ref{KGNoether} is a generalization of the result of \cite{Bozhkov}
for the Klein Gordon equation with constant potential, i.e. $V\left(
x^{k}\right) =V_{0}$. \ Furthermore, from Theorem \ref{KGNoether} we have
the following corollary which relates the dimension of the algebras of Lie
and Noether point symmetries of the Klein Gordon equation.

\begin{corollary}
If the Klein Gordon equation (\ref{KG.Eq1}) with Lagrangian (\ref{KGL.01})
admits $n$ Lie point symmetries which span the Lie algebra $G_{LS},$ $\dim
G_{LS}=n$, then the Lagrangian (\ref{KGL.01}) admits as Noether point
symmetries the Lie algebra $G_{NS}$, of dimension $\dim G_{NS}=n-1$, where $%
G_{LS}=G_{NS}\cup \left\{ u\partial _{u}\right\} .$
\end{corollary}

In the following we apply theorems \ref{KG} and \ref{KGNoether} in order to
determine the potentials for which the Klein Gordon equation (\ref{KG.Eq1})
admits Lie/Noether point symmetries in (diagonal) Bianchi I spacetimes.

\section{The Lie and Noether point symmetries of the Klein Gordon equation
in Bianchi I spacetime}

\label{KGBianchi}

In order to apply the results of theorems \ref{KG} and \ref{KGNoether} we
need the complete conformal algebra of Bianchi I\ spacetime.

\subsection{The conformal algebra of the diagonal Bianchi I\ spacetime}

The Bianchi type $N$ ($N=I,...,IX$) models are spatially homogeneous
spacetimes which admit a group of motions $G_{3}$ \cite{BRyan} acting on
spacelike hypersurfaces. Some \ of these spacetimes are non-isotropic
generalizations of the Friedman-Robertson-Walker (FRW) space-time such as
the Bianchi I, V and IX spacetimes. These spacetimes have been used in the
discussion of anisotropies in a primordial universe and its evolution
towards the observed isotropy of the present epoch.

The simplest type of these spacetimes are the Bianchi I models which
correspond to the abelian group $G_{3}$ consisting of the three KVs
\begin{equation}
Y_{I}^{1}=\partial _{x}~,~Y_{I}^{2}=\partial _{y}~,~Y_{I}^{3}=\partial _{z}.
\label{KGB.03a}
\end{equation}%
In synchronous coordinates the metric of this spacetime is
\begin{equation}
ds%
{{}^2}%
=-dt+A%
{{}^2}%
(t)dx+B%
{{}^2}%
(t)dy+C%
{{}^2}%
(t)dz  \label{BKG.01}
\end{equation}%
where $A(t),B(t),C(t)$ are functions of the time coordinate $t$ only.

When the metric functions $A(t),B(t),C(t)$ satisfy certain relations it is
possible that the resulting Bianchi I spacetimes admit a larger conformal
algebra. In \cite{TsAp} it has been shown that these spacetimes can be
classified in two sets a. The Bianchi I\ spacetimes which are
non-conformally flat (which we refer as Class A) and b. The Bianchi I
spacetimes which are conformally flat (which we refer as Class B). In more
formal terms the first family contains the Petrov type I\ spacetimes and the
second family the Petrov type D spacetimes.

Before we proceed it will be useful if we clarify the following point. From
theorem \ref{KG} we have that in order the Klein Gordon equation (\ref%
{BKG.03}) \ to admits Lie /Noether point symmetries in a Riemannian space
\emph{two} conditions must be satisfied:

a. The metric has to admit a non-void conformal algebra, which is always the
case because all Bianchi I metrics admit the three KVs $Y^{\mu }_{I}.$

b. The potential $V\left( t,x,y,z\right) $ must satisfy the constraint
condition (\ref{KGT.02}) or (\ref{KGT.04}) depending on the dimension of the
space i.e. $n>2$ or $n=2$ respectively.

This means that for each CKV of the conformal algebra of Bianchi I\
spacetimes given in \cite{TsAp} we must solve condition (\ref{KGT.02})
(because $n=4$) and find those potentials for which it is satisfied. In
other words the constraint condition (\ref{KGT.02}) acts as a double
selection rule selecting for each CKV a corresponding potential or, if this
is not possible, abandoning the CKV for being a Lie point symmetry of the
Klein Gordon equation.

\subsection{The general Bianchi I spacetime}

The generic line element of Bianchi I spacetime is (\ref{BKG.01}) with $%
A(t), B(t), C(t)$ being general smooth functions of $t$ and admits the
abelian group of isometries $G_{3}$ consisting of the vector fields $%
Y_{I}^{1},Y_{I}^{2},Y_{I}^{3}$ (see equation (\ref{KGB.03a}))

For the line element (\ref{BKG.01}) the Lagrangian (\ref{KGL.01}) becomes%
\begin{equation}
L\left( x^{i},u,u_{,i}\right) =\frac{1}{2}ABC\left(
-u_{,t}^{2}+A^{-2}u_{,x}^{2}+B^{-2}u_{,y}^{2}+C^{-2}u_{,z}^{2}\right) -\frac{%
1}{2}ABCV\left( t,x,y,z\right) u^{2}
\end{equation}%
and the Klein Gordon equation (\ref{KG.Eq1}) is
\begin{equation}
-u_{,tt}+A^{-2}u_{,xx}+B^{-2}u_{,yy}+C^{-2}u_{,zz}-\left( \frac{\dot{A}}{A}+%
\frac{\dot{B}}{B}+\frac{\dot{C}}{C}\right) u_{,t}+V\left( t,x,y,z\right) u=0.
\label{BKG.03}
\end{equation}

For each of the vectors $Y_{I}^{1-3}$ and their linear combinations we
solved condition (\ref{KGT.02}) and found the potentials $V\left(
t,x,y,z\right) $ of table \ref{BianchiIG}.

\begin{table}[tbp] \centering%
\caption{Point symmetries and potentials for the Klein Gordon equation in
the Bianchi I spacetime for arbitrary functions $A(t), B(t)$ and $C(t)$}%
\begin{tabular}{ccc}
\hline\hline
\textbf{Potential} & \textbf{Lie Sym.} & \textbf{Noether Sym.} \\ \hline
$V\left( t,x,y,z\right) $ & $X_{u}=\partial _{u}$ & No \\
$V\left( t,y,z\right) $ & $Y_{I}^{1}$ & Yes \\
$V\left( t,x,z\right) $ & $Y_{I}^{2}$ & Yes \\
$V\left( t,x,y\right) $ & $Y_{I}^{3}$ & Yes \\
$V\left( t,y-\frac{b}{a}x,z\right) $ & $aY_{I}^{1}+bY_{I}^{2}$ & Yes \\
$V\left( t,z-\frac{b}{a}x,y\right) $ & $aY_{I}^{1}+bY_{I}^{3}$ & Yes \\
$V\left( t,x,z-\frac{b}{a}y\right) $ & $aY_{I}^{2}+bY_{I}^{3}$ & Yes \\
$V\left( t,y-\frac{b}{a}x,z-\frac{c}{a}x\right) $ & $%
aY_{I}^{1}+bY_{I}^{2}+cY_{I}^{3}$ & Yes \\ \hline\hline
\end{tabular}%
\label{BianchiIG}%
\end{table}%

Having considered the general case we continue with the special cases of
Class A\ and Class B of Bianchi I\ spacetimes.

\subsection{Class A: Bianchi Spacetimes}

In the special case where $B^{2}\left( t\right) =C^{2}\left( t\right) $ and $%
A^{2}\left( t\right) \neq B^{2}\left( t\right) ~$the spacetime (\ref{BKG.01}%
) admits the extra KV $Y_{I}^{4}=z\partial _{y}-y\partial _{z}~$\footnote{%
Similarly if $A^{2}\left( t\right) =B^{2}\left( t\right) ,~A^{2}\left(
t\right) \neq C^{2}\left( t\right) $ or $A^{2}\left( t\right) =C^{2}\left(
t\right) ,~A^{2}\left( t\right) \neq B^{2}\left( t\right) $ the spacetime (%
\ref{BKG.01}) admits a four dimension Killing algebra where the
corresponding extra KVs are $y\partial _{x}-x\partial _{y}$ or $z\partial
_{x}-x\partial _{z}~$respectively. These cases being similar we restrict our
study to the case where $B^{2}\left( t\right) =C^{2}\left( t\right) $.}.

The vector field $Y_{I}^{4}$ and all linear combinations with the vector
fields $Y_{I}^{1-3}$ when introduced in the condition (\ref{KGT.02}) give as
solutions the results of table \ref{BianchiIG4}.

\begin{table}[tbp] \centering%
\caption{Point symmetries and potentials for the Klein Gordon equation in
the Bianchi I spacetime with $A(t)\neq B(t) = C(t)$}%
\begin{tabular}{ccc}
\hline\hline
\textbf{Potential~} & \textbf{Lie Sym.} & \textbf{Noether Sym.} \\ \hline
$V\left( t,x,y^{2}+z^{2}\right) $ & $Y_{I}^{4}$ & Yes \\
$V\left( t,x-\frac{a}{b}\arctan \frac{y}{z},y^{2}+z^{2}\right) $ & $%
aY_{I}^{1}+bY_{I}^{4}$ & Yes \\
$V\left( t,x,\frac{1}{2}\left( y^{2}+z^{2}\right) +\frac{a}{b}z\right) $ & $%
aY_{I}^{2}+bY_{I}^{4}$ & Yes \\
$V\left( t,x,\frac{1}{2}\left( y^{2}+z^{2}\right) +\frac{a}{b}y\right) $ & $%
aY_{I}^{3}+bY_{I}^{4}$ & Yes \\
$V\left( t,x-\frac{a}{c}\arctan \frac{cy}{b+cz},\frac{1}{2}\left(
y^{2}+z^{2}\right) +\frac{b}{c}z\right) $ & $%
aY_{I}^{1}+bY_{I}^{2}+cY_{I}^{4} $ & Yes \\
$V\left( t,x-\frac{a}{c}\arctan \frac{cy-b}{cz},\frac{1}{2}\left(
y^{2}+z^{2}\right) +\frac{b}{c}y\right) $ & $%
aY_{I}^{1}+bY_{I}^{3}+cY_{I}^{4} $ & Yes \\
$V\left( t,x,\frac{c}{2}\left( y^{2}+z^{2}\right) -\left( by-az\right)
\right) $ & $aY_{I}^{2}+bY_{I}^{3}+cY_{I}^{4}$ & Yes \\
$V\left( t,x-\frac{a}{d}\arctan \left( \frac{d~y-c}{d~z+c}\right) ,\frac{d}{2%
}\left( y^{2}+z^{2}\right) -\left( cy-bz\right) \right) $ & $%
aY_{I}^{1}+bY_{I}^{2}+cY_{I}^{3}+dY_{I}^{4}$ & Yes \\ \hline\hline
\end{tabular}%
\label{BianchiIG4}%
\end{table}%

When the functions $A\left( t\right) ,B\left( t\right) $ and $C\left(
t\right) $ satisfy the relations
\begin{equation}
A\left( t\right) =\frac{1}{U\left( t\right) }\bar{A}\left( t\right)
~,~B\left( t\right) =\frac{1}{U\left( t\right) }\bar{B}\left( t\right)
~,~C\left( t\right) =\frac{1}{U\left( t\right) }\bar{C}\left( t\right)
\label{BKG.06}
\end{equation}%
where
\begin{equation}
\bar{A}\left( t\right) =e^{-\alpha \int U\left( t\right) dt}~,~\bar{B}\left(
t\right) =e^{-\beta \int U\left( t\right) dt}~,~\bar{C}\left( t\right)
=e^{-\gamma \int U\left( t\right) dt}  \label{BKG.06a}
\end{equation}%
then the Bianchi I spacetime admits the proper CKV
\[
Y_{I}^{5}=\frac{1}{U\left( t\right) }\partial _{t}+\alpha x\partial
_{x}+\beta y\partial _{y}+\gamma z\partial _{z}
\]%
with conformal factor $\psi _{I}^{5}=-\frac{\dot{U}}{U^{2}}$~\cite{TsAp}. We
study two cases (a) the conformal factor $\psi _{I}^{5}$ is a solution of
Laplace equation $\Delta \psi _{I}^{5}=0$ and (b) the conformal factor $\psi
_{I}^{5}$ is not a solution of Laplace equation.

In case (a), the conformal factor $\psi _{I}^{5}$ is \ a solution of the
Laplace equation $\Delta \psi _{I}^{5}=0$ if the function $U\left( t\right) $
satisfies the condition
\begin{equation}
L^{\left( 3\right) }+\left( 3L^{\left( 1\right) }-\alpha -\beta -\gamma
\right) \frac{L^{\left( 2\right) }}{L}=0~,~\text{where }L=\frac{1}{U\left(
t\right) }.  \label{BKG.05}
\end{equation}

In order to find an exact solution of equation (\ref{BKG.05}) we use the Lie
symmetry method. By applying the Lie symmetry condition (\ref{pr.04}) where
the generator of the infinitesimal transformation is $\mathbf{X}_{L}=\xi
\left( t,L\right) \partial _{t}+\eta \left( t,L\right) \partial _{L}$, we
find that equation (\ref{BKG.05}) admits the extra Lie point symmetries $%
\partial _{t},t\partial _{t}+L\partial _{L}$. Therefore from the application
of invariants of Lie point symmetries in equation (\ref{BKG.05}), we find
the solutions%
\begin{equation}
L_{1}\left( t\right) =\frac{1}{M}W\left( \exp \left( M^{2}\left( \alpha
,\beta ,\gamma \right) \left( t+c_{1}\right) \right) \right) +\frac{1}{%
M\left( \alpha ,\beta ,\gamma \right) }~,~L_{2}\left( t\right) =t
\label{BKG.07}
\end{equation}%
where $M\left( \alpha ,\beta ,\gamma \right) =\alpha +\beta +\gamma $ and $%
W\left( t\right) $ is the Lambert W-function.

We observe that for the solution $L_{2}\left( t\right) ~$of (\ref{BKG.05}),
the conformal factor $\psi _{I}^{5}$ becomes a constant; that is, the CKV $%
Y_{I}^{5}$ reduces to a HV. This is the only Bianchi I spacetime which
admits a proper HV. In table \ref{BianchiIGCKV} we give the potentials which
admit a Lie / Noether point symmetry which is generated by the vector field $%
Y_{I}^{5}~$and all linear combinations of the vector field $Y_{I}^{5}$ with
the KVs $Y_{I}^{1-3}$.

\begin{table}[tbp] \centering%
\caption{Point symmetries and potentials for the Klein Gordon equation in
the Bianchi I spacetime with $A(t), B(t), C(t)$ are given from
(\ref{BKG.06}) and $U(t)$ from (\ref{BKG.07})}%
\begin{tabular}{ccc}
\hline\hline
\textbf{Potential~}$V\left( t,x,y,z\right) =U^{2}\left( t\right) \bar{V}%
\left( t,x,y,z\right) $ & \textbf{Lie Sym.} & \textbf{Noether Sym.} \\ \hline
$\bar{V}\left( x\bar{A},y\bar{B},z\bar{C}\right) $ & $Y_{I}^{5}-\psi
_{I}^{5}u\partial _{u}$ & Yes \\
$\bar{V}\left( \left( x+\frac{1}{\alpha }\frac{a}{b}\right) \bar{A},y\bar{B}%
,z\bar{C}\right) $ & $aY_{I}^{1}+bY_{I}^{5}-b\psi _{I}^{5}u\partial _{u}$ &
Yes \\
$\bar{V}\left( x\bar{A},\left( y+\frac{1}{\beta }\frac{a}{b}\right) y\bar{B}%
,z\bar{C}\right) $ & $aY_{I}^{2}+bY_{I}^{5}-b\psi _{I}^{5}u\partial _{u}$ &
Yes \\
$\bar{V}\left( x\bar{A},y\bar{B},\left( z+\frac{1}{\gamma }\frac{a}{b}%
\right) \bar{C}\right) $ & $aY_{I}^{3}+bY_{I}^{5}-b\psi _{I}^{5}u\partial
_{u}$ & Yes \\
$\bar{V}\left( \left( x+\frac{1}{\alpha }\frac{a}{c}\right) \bar{A},\left( y+%
\frac{1}{\beta }\frac{b}{c}\right) \bar{B},z\bar{C}\right) $ & $%
aY_{I}^{1}+bY_{I}^{2}+cY_{I}^{5}-c\psi _{I}^{5}u\partial _{u}$ & Yes \\
$\bar{V}\left( \left( x+\frac{1}{\alpha }\frac{a}{c}\right) \bar{A},y\bar{B}%
,\left( z+\frac{1}{\gamma }\frac{b}{c}\right) \bar{C}\right) $ & $%
aY_{I}^{1}+bY_{I}^{3}+cY_{I}^{5}-c\psi _{I}^{5}u\partial _{u}$ & Yes \\
$\bar{V}\left( x\bar{A},\left( y+\frac{1}{b}\frac{a}{c}\right) \bar{B}%
,\left( z+\frac{1}{\gamma }\frac{b}{c}\right) \bar{C}\right) $ & $%
aY_{I}^{2}+bY_{I}^{3}+cY_{I}^{5}-c\psi _{I}^{5}u\partial _{u}$ & Yes \\
$\bar{V}\left( \left( x+\frac{1}{\alpha }\frac{a}{d}\right) \bar{A},\left( y+%
\frac{1}{\beta }\frac{b}{d}\right) \bar{B},\left( z+\frac{1}{\gamma }\frac{c%
}{d}\right) \bar{C}\right) $ & $%
aY_{I}^{1}+bY_{I}^{2}+cY_{I}^{3}+dY_{I}^{5}-d\psi _{I}^{5}u\partial _{u}$ &
Yes \\ \hline\hline
\end{tabular}%
\label{BianchiIGCKV}%
\end{table}%

In case (b), where $U\left( t\right) $ is arbitrary, the Klein Gordon
equation (\ref{BKG.03}) admits the Lie/ Noether point symmetries of table %
\ref{BianchiIGCKV} when
\begin{equation}
V\left( t,x,y,z\right) =U^{2}\bar{V}\left( t,x,y,z\right) -\frac{1}{U}%
U^{\left( 2\right) }+\left( \frac{3U^{\left( 1\right) }}{U^{2}}+M\left(
\alpha ,\beta ,\gamma \right) \right) U^{\left( 1\right) }\text{. }
\end{equation}%
and $\bar{V}\left( t,x,y,z\right) $ are taken from table \ref{BianchiIGCKV}.

Furthermore when $\bar{B}\left( t\right) =\bar{C}\left( t\right) ,~$i.e. $%
\beta =\gamma $ in (\ref{BKG.06a}), the spacetime (\ref{BKG.01}) admits also
the additional KV $Y_{I}^{4}$. Then the general (extra) point symmetry
vector is~$X_{GI}=\sum_{\alpha =1}^{4}a_{\alpha }Y_{I}^{\alpha
}+a_{5}X_{I}^{5}$ where $X_{I}^{5}=Y_{I}^{5}-\psi _{I}^{5}u\partial _{u}$.
In table \ref{BianchiICom1} we give the commutators of the elements of the
generic point symmetry $X_{GI}.$ These will be used in the reduction of the
Klein Gordon equation.

\begin{table}[tbp] \centering%
\caption{Commutators of the elements of the generic Lie symmetry vector of
the Klein Gordon equation in a Bianchi I spacetime where $A(t), B(t) = C(t)$
are given from (\ref{BKG.06}) and $U(t)$ from (\ref{BKG.07})}%
\begin{tabular}{c|ccccc}
\hline\hline
$\left[ .,.\right] $ & $Y_{I}^{1}$ & $Y_{I}^{2}$ & $Y_{I}^{3}$ & $Y_{I}^{4}$
& $X_{I}^{5}$ \\ \hline
$Y_{I}^{1}$ & $0$ & $0$ & $0$ & $0$ & $\alpha Y_{I}^{1}$ \\
$Y_{I}^{2}$ &  & $0$ & $0$ & $-Y_{I}^{3}$ & $\beta Y_{I}^{2}$ \\
$Y_{I}^{3}$ &  &  & $0$ & $Y_{I}^{2}$ & $\beta Y_{I}^{3}$ \\
$Y_{I}^{4}$ &  &  &  & $0$ & $0$ \\
$X_{I}^{5}$ &  &  &  &  & $0$ \\ \hline\hline
\end{tabular}%
\label{BianchiICom1}%
\end{table}%

\subsection{Class B: Bianchi I spacetime is conformally flat}

According to \cite{TsAp} the line element of the conformally flat Bianchi I
spacetime (\ref{BKG.01}) (excluding the FRW spacetime) is of the following
two forms%
\begin{equation}
ds^{2}=-dt^{2}+\sin ^{2}tdx^{2}+\cos ^{2}tdy^{2}+dz^{2}  \label{BKG.08}
\end{equation}%
and%
\begin{equation}
ds^{2}=-dt^{2}+\sinh ^{2}tdx^{2}+\cosh ^{2}tdy^{2}+dz^{2}.  \label{BKG.09}
\end{equation}%
The conformal algebra of these spaces consists of 15 vector fields, in which
the seven vector fields are KVs and the remaining eight vector fields are
proper CKVs \cite{TsAp}.

The corresponding Lagrangian of the Klein Gordon equation in each space is:%
\begin{equation}
L_{1}=\frac{\sin 2t}{4}\left( -u_{,t}^{2}+u_{,z}^{2}+\left( \sin t\right)
^{-2}1u_{,x}+\left( \cos t\right) ^{-2}u_{,y}^{2}\right) -\frac{\sin 2t}{4}%
V\left( t,x,y,z\right) u^{2}  \label{BKG.10A}
\end{equation}%
and%
\begin{equation}
L_{2}=\frac{\sinh 2t}{4}\left( -u_{,t}^{2}+u_{,z}^{2}+\left( \sinh t\right)
^{-2}u_{,x}+\left( \cosh t\right) ^{-2}u_{,y}^{2}\right) -\frac{\sinh 2t}{4}%
V\left( t,x,y,z\right) u^{2}  \label{BKG.11A}
\end{equation}

The Euler-Lagrange equation of (\ref{BKG.10A}) is the Klein Gordon equation
\begin{equation}
-u_{,tt}+\sin ^{-2}t~u_{,xx}+\cos ^{-2}t~u_{,yy}+u_{,zz}+\left( \tan t-\cot
t\right) u_{,t}+V\left( t,x,y,z\right) u=0  \label{BKG.10}
\end{equation}%
and the Euler-Lagrange equation of (\ref{BKG.11A}) gives%
\begin{equation}
-u_{,tt}+\sinh ^{-2}t~u_{,xx}+\cosh ^{-2}t~u_{,yy}+u_{,zz}-\left( \tanh
t+\coth t\right) u_{,t}+V\left( t,x,y,z\right) u=0  \label{BKG.11}
\end{equation}

The Klein Gordon equations (\ref{BKG.10}) and (\ref{BKG.11}) are related by
the complex coordinate transformation $\left( t,x,y,z\right) \rightarrow
\left( i\bar{t},i\bar{x},i\bar{y},i\bar{z}\right) $, therefore in the
following we study only the Lie point symmetries of equation (\ref{BKG.10}).

For the metric (\ref{BKG.08}) the seven KVs are the three vector fields $%
Y_{I}^{1-3}$ and the four vector fields%
\[
\bar{Y}_{I}^{4}=e^{x+y}\left( \partial _{t}-\cot t\partial _{x}+\tan
t\partial _{y}\right) \ ,~\bar{Y}_{I}^{5}=e^{x-y}\left( \partial _{t}-\cot
t\partial _{x}-\tan t\partial _{y}\right)
\]%
\[
\bar{Y}_{I}^{6}=e^{-x+y}\left( \partial _{t}+\cot t\partial _{x}+\tan
t\partial _{y}\right) ~,~\bar{Y}^{7}=e^{-x-y}\left( \partial _{t}+\cot
t\partial _{x}-\tan t\partial _{y}\right) .
\]%
Concerning the eight CKVs and their corresponding conformal factors these
are:%
\[
C_{\pm x}^{1}=e^{\pm x}\left( \cos t\cos z\partial _{t}\mp \frac{\cos z}{%
\sin t}\partial _{x}-\sin t\sin z\partial _{z}\right) ,~\psi _{\pm
x}^{1}=-e^{\pm x}\sin t\cos z
\]%
\[
C_{\pm x}^{2}=e^{\pm x}\left( \cos t\sin z\partial _{t}\mp \frac{\sin z}{%
\sin t}\partial _{x}+\sin t\cos z\partial _{z}\right) ,~\psi _{\pm
x}^{2}=-e^{\pm x}\sin t\sin z
\]%
\[
C_{\pm y}^{1}=e^{\pm y}\left( \sin t\sin z\partial _{t}\pm \frac{\sin z}{%
\cos t}\partial _{y}-\cos t\cos z\partial _{z}\right) ,~\psi _{\pm
y}^{1}=e^{\pm y}\cos t\sin z
\]%
\[
C_{\pm y}^{2}=e^{\pm y}\left( \sin t\cos z\partial _{t}\pm \frac{\cos z}{%
\cos t}\partial _{y}+\cos t\sin z\partial _{z}\right) ,~\psi _{\pm
y}^{2}=e^{\pm y}\cos t\cos z
\]

Having the complete conformal algebra of the space we continue with the
application of theorems \ref{KG} and \ref{KGNoether} in order to determine
the potentials for which the Klein Gordon equation (\ref{BKG.10}) admits Lie
and Noether point symmetries. For convenience and economy of space the
results are presented in tabular form. In table \ref{BianchiIGCF} we give
the form of the potentials for which the Lie and the Noether point
symmetries are generated by the Killing subalgebra $\left\{ \bar{Y}%
^{4-7}\right\} ~$of the metric (\ref{BKG.08}) and in table \ref{BianchiIGCF2}
we give the potentials for the point symmetries which are generated by the
proper Conformal algebra $\left\{ C_{\pm \left( x,y\right) }^{1-2}\right\} $
of the metric (\ref{BKG.08}).

\begin{table}[tbp] \centering%
\caption{Point symmetries and potentials for the Klein
Gordon equation in the Bianchi I spacetime with metric (\ref{BKG.08})}%
\begin{tabular}{ccc}
\hline\hline
\textbf{Potential} & \textbf{Lie Sym.} & \textbf{Noether Sym.} \\ \hline
$V\left( x+\ln \left( \sin t\right) ,y+\ln \left( \cos t\right) ,z\right) $
& $\bar{Y}_{I}^{4}$ & Yes \\
$V\left( x+\ln \left( \sin t\right) ,y-\ln \left( \cos t\right) ,z\right) $
& $\bar{Y}_{I}^{5}$ & Yes \\
$V\left( x-\ln \left( \sin t\right) ,y+\ln \left( \cos t\right) ,z\right) $
& $\bar{Y}_{I}^{6}$ & Yes \\
$V\left( x-\ln \left( \sin t\right) ,y-\ln \left( \cos t\right) ,z\right) $
& $\bar{Y}_{I}^{7}$ & Yes \\ \hline\hline
\end{tabular}%
\label{BianchiIGCF}%
\end{table}%

\begin{table}[tbp] \centering%
\caption{Point symmetries and potentials for the Klein
Gordon equation in the Bianchi I spacetime with metric (\ref{BKG.08})}%
\begin{tabular}{ccc}
\hline\hline
\textbf{Potential} & \textbf{Lie Sym.} & \textbf{Noether Sym.} \\ \hline
$1-\frac{1}{2\cos ^{2}t}+\frac{1}{\cos ^{2}t}V\left( x\pm \ln \left( \frac{%
1-\cos 2t}{\sin 2t}\right) ,y,\frac{\cos t}{\sin z}\right) $ & $C_{\pm
x}^{1}-\psi _{\pm x}^{1}u\partial _{u}$ & Yes \\
$1-\frac{1}{2\cos ^{2}t}+\frac{1}{\cos ^{2}t}V\left( x\pm \ln \left( \frac{%
1-\cos 2t}{\sin 2t}\right) ,y,\frac{\cos t}{\cos z}\right) $ & $C_{\pm
x}^{2}-\psi _{\pm x}^{2}u\partial _{u}$ & Yes \\
$1-\frac{1}{2\sin ^{2}t}+\frac{1}{\sin ^{2}t}V\left( x,y\mp \left( \frac{%
1-\cos 2t}{\sin 2t}\right) ,\frac{\sin t}{\cos z}\right) $ & $C_{\pm
y}^{1}-\psi _{\pm y}^{1}u\partial _{u}$ & Yes \\
$1-\frac{1}{2\sin ^{2}t}+\frac{1}{\sin ^{2}t}V\left( x,y\mp \left( \frac{%
1-\cos 2t}{\sin 2t}\right) ,\frac{\sin t}{\sin z}\right) $ & $C_{\pm
y}^{2}-\psi _{\pm y}^{2}u\partial _{u}$ & Yes \\ \hline\hline
\end{tabular}%
\label{BianchiIGCF2}%
\end{table}%

\section{Lie invariant solutions of the Klein Gordon equation in Bianchi I
spacetimes}

\label{invariant}

In this section we apply the zero order invariants of some Lie point
symmetries of the Klein Gordon equation (\ref{BKG.03}) in order to reduce
the differential equation and determine invariant solutions with respect to
these symmetries. We will study the reduction for the two class A and B of
section \ref{KGBianchi}

\subsection{Class A: Invariant solutions}

\label{casea1}

For the class A Bianchi I spacetimes we will study the reduction for the of
the Klein Gordon eqation for two cases, (I) when $V\left( t,x,y,z\right)
=V\left( t\right) $ and $A^{2}\left( t\right) \neq B^{2}\left( t\right) \neq
C^{2}\left( t\right) $ and (II) when $V\left( t,x,y,z\right) =V\left( \left(
y^{2}+z^{2}\right) e^{-2\beta \int U\left( t\right) dt}\right) U^{2}\left(
t\right) $ where $\,A\left( t\right) ,~B\left( t\right) ,~C\left( t\right) $
are given from (\ref{BKG.06}) and $U\left( t\right) =\frac{1}{t}$.

\paragraph{Case I:}

When $V=V\left( t\right) $ from table \ref{BianchiIG} the Klein Gordon
equation (\ref{BKG.03}) admits as extra Lie point symmetries the vector
fields $Y_{I}^{1-3}.$ However in order to apply the Lie invariants we choose
the symmetries%
\[
X_{I}^{\alpha }=Y_{I}^{\alpha }+\mu _{\alpha }X_{u}~,~\alpha =1,2,3
\]%
which are linear combinations with the symmetry $X_{u}.~$Furthermore, for
the commutators we have $\left[ X_{I}^{a},X_{I}^{\beta }\right] =0$, where $%
\alpha ,\beta =1,2,3$. The zero order invariants of $X_{I}^{1}$ are $\left\{
t,y,z,e^{-\mu _{1}x}u\right\} $. We select $\left\{ t,y,z\right\} $ to be
the independent variables and $u\left( t,x,y,z\right) =e^{\mu _{1}x}v\left(
t,y,z\right) $, where $v\left( t,y,z\right) $ is the dependent variable. By
replacing in (\ref{BKG.03}) we find the reduced equation%
\begin{equation}
-v_{,tt}+B^{-2}v_{,yy}+C^{-2}v_{,zz}-\left( \frac{\dot{A}}{A}+\frac{\dot{B}}{%
B}+\frac{\dot{C}}{C}\right) v_{,t}+\left( V\left( t\right) +\mu
_{1}^{2}\right) v=0.  \label{IBKG.04}
\end{equation}%
The Lie point symmetries $X_{I}^{2,3}$ are inherited symmetries of (\ref%
{IBKG.04}); therefore, by applying the zero order invariants the solution of
the Klein Gordon equation (\ref{BKG.03}) is
\begin{equation}
u\left( t,x,y,z\right) =\exp \left( \mu _{1}x+\mu _{2}y+\mu _{3}z\right)
w\left( t\right)  \label{IBKG.05}
\end{equation}%
where $w\left( t\right) $ is the solution of the second order ODE%
\begin{equation}
\ddot{w}+\left( \frac{\dot{A}}{A}+\frac{\dot{B}}{B}+\frac{\dot{C}}{C}\right)
\dot{w}-\left( V\left( t\right) +\mu _{1}^{2}+\mu _{2}^{2}+\mu
_{3}^{2}\right) w=0.  \label{IBKG.06}
\end{equation}%
This is the damped oscillator which admits eight Lie point symmetries
therefore can be transformed to the equation $W^{\prime \prime }+\omega
^{2}W=0$ by an appropriate transformation~\cite{GovinderLeach}.

\paragraph{Case II:}

In the second case we consider the potential $V\left( t,x,y,z\right) =\frac{1%
}{t^{2}}V^{\prime }\left( \frac{y^{2}+z^{2}}{t^{\beta }}\right) $ and we
select the line element of the Bianchi I spacetime to be%
\begin{equation}
ds^{2}=-dt^{2}+t^{2-2\alpha }dx^{2}+t^{2-2\beta }\left( dy^{2}+dz^{2}\right)
\label{IBKG.07}
\end{equation}%
hence the corresponding Klein Gordon equation is
\begin{equation}
-u_{,tt}+t^{2\alpha -2}u_{,xx}+t^{2\beta -2}\left( u_{,yy}+u_{,zz}\right) -%
\frac{3-\alpha -2\beta }{t}u_{,t}+\frac{1}{t^{2}}V^{\prime }\left( \frac{%
y^{2}+z^{2}}{t^{2\beta }}\right) u=0  \label{IBKG.08}
\end{equation}

From tables \ref{BianchiIG}, \ref{BianchiIG4} and \ref{BianchiIGCKV}, we
have that the Klein Gordon equation (\ref{IBKG.08}) admits three extra Lie
point symmetries, the $Y_{I}^{1},Y_{I}^{4}$ and the $X_{I}^{5}=Y_{I}^{5}-u%
\partial _{u}$, where $\psi _{I}^{5}=1$~since for the metric (\ref{IBKG.07})
$Y_{I}^{5}$ is a HV. In order to apply the zero order invariants we select
the symmetries%
\[
X_{I}^{1}=Y_{I}^{1}+\mu _{1}X_{u}~,~X_{I}^{4}=Y_{I}^{4}+\mu
_{2}X_{u}~,~X_{I}^{5^{\prime }}=X_{I}^{5}+\mu _{5}X_{u}
\]%
with commutators%
\begin{equation}
\left[ X_{I}^{1},Y_{I}^{4}\right] =0,\left[ X_{I}^{5^{\prime }},X_{I}^{4}%
\right] =0~,~\left[ X_{I}^{1},X_{I}^{5^{\prime }}\right] =\alpha Y_{I}^{1}.
\label{IBKG.09A}
\end{equation}

We start the reduction with the vector field $X_{I}^{4}$. The zero order
invariants are $\left\{ t,x,y^{2}+z^{2},e^{-\mu _{4}\arctan \frac{z}{y}%
}u\right\} .$ We select $t,x$ and $r=x^{2}+y^{2}$ to be the independent
variables and $u\left( t,x,y,z\right) =e^{\mu _{4}\theta }v\left(
t,x,r\right) $, where $v\left( t,x,r\right) $ is the dependent variable and $%
\theta =\arctan \frac{z}{y}$. Replacing in (\ref{IBKG.08}) we find the
reduced equation%
\begin{equation}
-v_{,tt}+t^{2\alpha -2}v_{,xx}+t^{2\beta -2}\left( v\,_{,rr}+\frac{1}{r}%
v_{,r}\right) -\frac{3-\alpha -2\beta }{t}v_{,t}+\frac{1}{t^{2}}\bar{V}%
\left( \frac{r^{2}}{t^{2\beta }}\right) v=0  \label{IBKG.09}
\end{equation}%
where $\bar{V}\left( \frac{r^{2}}{t^{2\beta }}\right) =V^{\prime }\left(
\frac{r^{2}}{t^{2\beta }}\right) +\mu _{4}\frac{t^{2\beta }}{r^{2}}$.
Equation (\ref{IBKG.09}) admits as Lie point symmetries the vector fields $%
X_{I}^{1},X_{I}^{5^{\prime }}$ which are inherited symmetries. \ We continue
with the application of the zero order invariants of $X_{I}^{1}$. These are $%
\left\{ t,r,e^{-\mu _{1}x}v\right\} $, hence we select as independent
variables the $t,r$ and as dependent variable the $w\left( t,r\right)
=e^{-\mu _{1}x}v$. By replacing in~(\ref{IBKG.09}) we have%
\begin{equation}
-w_{,tt}+t^{2\beta -2}\left( w\,_{,rr}+\frac{1}{r}w_{,r}\right) -\frac{%
3-\alpha -2\beta }{t}w_{,t}+\frac{1}{t^{2}}\left( \bar{V}\left( \frac{r^{2}}{%
t^{2\beta }}\right) +\mu _{1}^{2}t^{2\alpha }\right) w=0  \label{IBKG.10}
\end{equation}

However, from the commutator (\ref{IBKG.09A}) the Lie point symmetry $%
X_{I}^{5^{\prime }}$ is an inherited symmetry when $X_{I}^{1}=Y_{I}^{1}$,
i.e. $\mu _{1}=0$, or when $\alpha =0$. \ From the vector field $%
X_{I}^{5^{\prime }}$ we have that the zero order invariants are $\zeta
=rt^{-\beta }$ and $w\left( t,r\right) =t^{\mu _{5}}\sigma \left( \zeta
\right) $. Therefore the reduced equation is%
\begin{equation}
\left( 1-\zeta ^{2}\beta ^{2}\right) \sigma _{,\zeta \zeta }-\beta \zeta
\left( 3\beta -2-2\mu _{5}-\frac{1}{\beta \zeta }\right) \sigma _{,\zeta
}+K\left( \zeta \right) \sigma =0.  \label{IBKG.11}
\end{equation}%
When $\mu _{1}=0$, $K\left( \zeta \right) =2\beta \mu _{5}+\bar{V}\left(
\zeta \right) $ and the solution of the Klein Gordon equation (\ref{IBKG.08}%
) is
\begin{equation}
u\left( t,x,y,z\right) =t^{\mu _{5}}\exp \left( \mu _{4}\arctan \left( \frac{%
z}{y}\right) \right) \sigma \left( \sqrt{y^{2}+z^{2}}t^{-\beta }\right)
\end{equation}%
When $\alpha =0$,~$K\left( \zeta \right) =2\beta \mu _{5}+\mu _{1}^{2}+\bar{V%
}\left( \zeta \right) $ and the solution of the Klein Gordon equation (\ref%
{IBKG.08}) is%
\begin{equation}
u\left( t,x,y,z\right) =t^{\mu _{5}}\exp \left( \mu _{1}x+\mu _{4}\arctan
\left( \frac{z}{y}\right) \right) \sigma \left( \sqrt{y^{2}+z^{2}}t^{-\beta
}\right)
\end{equation}%
where $\sigma \left( \sqrt{y^{2}+z^{2}}t^{-\beta }\right) =\sigma \left(
\zeta \right) $ satisfies the second order ODE (\ref{IBKG.11}).

\subsection{Class B: Invariant solutions}

In this section, we use the Lie symmetries of the Klein Gordon equation (\ref%
{BKG.10}) in order to find invariant analytic solutions. We shall do that
for the point symmetries generated by the KVs only. We shall consider only
the potentials $V_{\pm x}=V\left( x\pm \ln \left( \sin t\right) \right) $
and $V_{\pm y}=V\left( y\pm \ln \left( \cos t\right) \right) $.

For the reduction of (\ref{BKG.10}) we need the commutators of the KVs of
the metric (\ref{BKG.08}) which are given in table \ref{BianchiICom}.

\begin{table}[tbp] \centering%
\caption{Commutators of the elements of the generic Lie point symmetry vector of
the Klein Gordon equation in a Bianchi I spacetime with metric
(\ref{BKG.08})}%
\begin{tabular}{c|ccccccc}
\hline\hline
$\left[ .,.\right] $ & $Y_{I}^{1}$ & $Y_{I}^{2}$ & $Y_{I}^{3}$ & $\bar{Y}%
_{I}^{4}$ & $\bar{Y}_{I}^{5}$ & $\bar{Y}_{I}^{6}$ & $\bar{Y}_{I}^{7}$ \\
\hline
$Y_{I}^{1}$ & $0$ & $0$ & $0$ & $\bar{Y}_{I}^{4}$ & $\bar{Y}_{I}^{5}$ & $-%
\bar{Y}_{I}^{6}$ & $-\bar{Y}_{I}^{7}$ \\
$Y_{I}^{2}$ &  & $0$ & $0$ & $\bar{Y}_{I}^{4}$ & $-\bar{Y}_{I}^{5}$ & $\bar{Y%
}_{I}^{6}$ & $-\bar{Y}_{I}^{7}$ \\
$Y_{I}^{3}$ &  &  & $0$ & $0$ & $0$ & $0$ & $0$ \\
$\bar{Y}_{I}^{4}$ &  &  &  & $0$ & $0$ & $0$ & $-4\left(
Y_{I}^{1}+Y_{I}^{2}\right) $ \\
$\bar{Y}_{I}^{5}$ &  &  &  &  & $0$ & $4\left( -Y_{I}^{1}+Y_{I}^{2}\right) $
& $0$ \\
$\bar{Y}_{I}^{6}$ &  &  &  &  &  & $0$ & $0$ \\
$\bar{Y}_{I}^{7}$ &  &  &  &  &  &  & $0$ \\ \hline\hline
\end{tabular}%
\label{BianchiICom}%
\end{table}%

\paragraph{Potential $V_{+x}=V\left( x+\ln \left( \sin t\right) \right) $}

From tables \ref{BianchiIG} and \ref{BianchiIGCF} we read that for this
potential the Klein Gordon equation (\ref{BKG.10}) admits four extra Lie
point symmetries given by the vector fields $Y_{I}^{2},Y_{I}^{3},\bar{Y}%
_{I}^{4},\bar{Y}_{I}^{5}.$

In order to reduce equation (\ref{BKG.10}) and determine an invariant
solution we need a double reduction of the equation, therefore we must have
a Lie point symmetry of the reduced equation. This is assured if we use Lie
point symmetries which commute hence reduction by any of them inherits the
remaining vector to the reduced equation. In our case we shall use for
reduction the Lie point symmetries
\[
X_{I}^{3}=Y_{I}^{3}+\mu _{3}X_{u},~~\bar{X}_{I}^{4}=\bar{Y}_{I}^{4}+\mu
_{4}X_{u}~,~\bar{X}_{I}^{5}=\bar{Y}_{I}^{5}+\mu _{5}X_{u}
\]%
of table \ref{BianchiIG} because from table \ref{BianchiICom} we have that
all their commutators vanish.

Reduction by $X_{I}^{3}$ leads to the reduced equation%
\begin{equation}
-v_{,tt}+\sin ^{-2}t~v_{,xx}+\cos ^{-2}t~v_{,yy}+\left( \tan t-\cot t\right)
v_{,t}+\left( V\left( x+\ln \left( \sin t\right) \right) +\mu
_{3}^{2}\right) v=0  \label{BKG.14}
\end{equation}%
where $u\left( t,x,y,z\right) =v\left( t,x,y\right) e^{\mu _{3}z}.$ As we
explained above, equation (\ref{BKG.14}) inherits the vector fields $\bar{X}%
_{I}^{4},\bar{X}_{I}^{5}$ as Lie point symmetries. Reduction of (\ref{BKG.14}%
) by $\bar{X}_{I}^{4},\bar{X}_{I}^{5}~$leads to the solution of the Klein
Gordon equation:%
\begin{equation}
u\left( t,x,y,z\right) =\sigma \left( \zeta \right) \exp (\mu _{3}z-\frac{%
\cot t}{2}\left( \mu _{4}e^{-\left( x+y\right) }+\mu _{5}e^{x-y}\right)
\end{equation}%
where$~\zeta =x+\ln \left( \sin t\right) $ and $\sigma \left( \zeta \right) $
satisfies the second order ODE%
\begin{equation}
\sigma _{,\zeta \zeta }+2\sigma _{,\zeta }+\left( V\left( \zeta \right) +\mu
_{3}^{2}-\mu _{4}\mu _{5}e^{-2\zeta }\right) \sigma =0.  \label{BKG.15}
\end{equation}

In the case where $V\left( \zeta \right) =e^{-2\zeta }\mu _{4}\mu _{5}$,
i.e.
\[
V\left( x+\ln \left( \sin t\right) \right) =\mu _{4}\mu _{5}\exp \left(
-\left( x+\ln \left( \sin t\right) \right) \right)
\]
the solution of (\ref{BKG.15}) is%
\begin{equation}
\sigma \left( \zeta \right) =\sigma _{1}\exp \left( \left[ -1+\sqrt{1-\mu
_{3}^{2}}\right] \zeta \right) +\sigma _{2}\exp \left( \left[ -1-\sqrt{1-\mu
_{3}^{2}}\right] \zeta \right) .  \label{BKG.16}
\end{equation}

\paragraph{Potential $V_{-x}=V\left( x-\ln \left( \sin t\right) \right) $}

When $V\left( t,x,y,z\right) =V\left( x-\ln \left( \sin t\right) \right) ,$
the Klein Gordon equation (\ref{BKG.10}) admits the extra Lie point
symmetries $Y_{I}^{2},Y_{I}^{3},\bar{Y}_{I}^{6},\bar{Y}_{I}^{7}$. With the
same reasoning as in the previous case we use for the reduction the KVs
\[
X_{I}^{3}=Y_{I}^{3}+\mu _{3}X_{u},~~\bar{X}_{I}^{6}=\bar{Y}_{I}^{6}+\mu
_{6}X_{u}~,~\bar{X}_{I}^{7}=\bar{Y}_{I}^{7}+\mu _{7}X_{u}
\]%
which span an Abelian algebra. Successive reduction of (\ref{BKG.10}) by $%
X_{I}^{3},\bar{X}_{I}^{6}$ and $\bar{X}_{I}^{7}$ results in the solution
\begin{equation}
u\left( t,x,y,z\right) =\sigma \left( \xi \right) \exp \left( \mu _{3}z-%
\frac{\cot t}{2}\left( \mu _{6}e^{x-y}+\mu _{7}e^{x+y}\right) \right) .
\label{BKG.18}
\end{equation}%
where $\xi =x-\ln \left( \sin t\right) $and $\sigma \left( \xi \right) $is a
solution of the equation:%
\begin{equation}
\sigma _{,\xi \xi }-2\sigma _{,\xi }+\left( V\left( \zeta \right) +\mu
_{3}^{2}-\mu _{6}\mu _{7}e^{2\xi }\right) \sigma =0  \label{BKG.17}
\end{equation}

Again we note that for the potential $~V_{-x}=\mu _{6}\mu _{7}\exp \left(
2\xi \right) -\mu _{3}^{2}$ the solution (\ref{BKG.18}) of the Klein Gordon
equation (\ref{BKG.10}) is:%
\begin{equation}
u\left( t,x,y,z\right) =\left[ \sigma _{0}+\sigma _{1}\exp \left( 2x-\ln
\left( \sin ^{2}t\right) \right) \right] \exp \left( \mu _{3}z-\frac{\cot t}{%
2}\left( \mu _{6}e^{x-y}+\mu _{7}e^{x+y}\right) \right)
\end{equation}

\paragraph{Potentials $V_{\pm y}:$}

When $V\left( t,x,y,z\right) =V\left( y+\ln \left( \cos t\right) \right) $,
the Klein Gordon equation (\ref{BKG.10}) admits the extra Lie point
symmetries~$Y_{I}^{1},Y_{I}^{3},\bar{Y}_{I}^{4},\bar{Y}_{I}^{6}.$ For the
reduction in this case we select the vectors $X_{I}^{3},~\bar{X}_{I}^{4},~%
\bar{X}_{I}^{6}~\ $and find the solution:%
\begin{equation}
u\left( t,x,y,z\right) =\sigma \left( \zeta \right) \exp \left( \mu _{3}z+%
\frac{\tan t}{2}\left( \mu _{4}e^{-x-y}+\mu _{6}e^{x-y}\right) \right)
\end{equation}%
where $\zeta =y+\ln \left( \cos t\right) $ and $\sigma \left( \zeta \right) $
satisfies equation (\ref{BKG.15}) with $\mu _{5}=\mu _{6}$.

When $V\left( t,x,y,z\right) =V\left( y-\ln \left( \cos t\right) \right) $,
the Klein Gordon equation (\ref{BKG.10}) admits the extra Lie point
symmetries~$Y_{I}^{1},Y_{I}^{3},\bar{Y}_{I}^{5},\bar{Y}_{I}^{7}.$ In this
case, we reduce equation (\ref{BKG.10}) with the Lie algebra $\left\{
X_{I}^{3},\bar{X}_{I}^{5},\bar{X}_{I}^{7}\right\} $ and we find the
following invariant solution%
\begin{equation}
u\left( t,x,y,z\right) =\sigma \left( \xi \right) \exp \left( \mu _{3}z+%
\frac{\tan t}{2}\left( \mu _{5}e^{-x+y}+\mu _{7}e^{x+y}\right) \right)
\end{equation}%
where $\xi =y-\ln \left( \cos t\right) $ and $\sigma \left( \xi \right) $
satisfies equation (\ref{BKG.17}) with $\mu _{6}=\mu _{5}$.

\section{Conclusion}

\label{conclusions}

The knowledge of the Lie point symmetries of a differential equation is
important because it can be used to determine invariant solutions of the
equation. In this work we considered the Klein Gordon (\ref{KG.Eq1}) in a
general Riemannian space and proved that (a) The Lie point symmetries of (%
\ref{KG.Eq1}) coincide with the Noether point symmetries of Lagrangian (\ref%
{KGL.01}) (b) The generators of the Lie/ Noether point symmetries are the
CKVs and their linear combinations (c) Not all CKVs of the space are Lie /
Noether point symmetries of (\ref{KG.Eq1}) ; A CKV is a Lie / Noether point
symmetry of (\ref{KG.Eq1}) if the constraint condition (\ref{KGT.02}) has a
solution for the corresponding potential.

This general results transfer the problem of determining the Lie and the
Noether point symmetries of the Klein Gordon equation in a general
Riemannian space to the problem of determining the CKVs of the space and the
solution of an easy differential condition.

We have applied the general results in the case of the Bianchi I spacetime.
The complete conformal algebra of the Bianchi I\ spacetimes has been
determined in \cite{TsAp}. We have used the results of \cite{TsAp} in order
a. To determine all potentials for which the resulting Klein Gordon equation
in Bianchi I\ spacetime admits Lie and Noether point symmetries and b. To
determined the Lie /Noether symmetry vectors. Due to the plethora of cases
and for easy reference the results are presented in the form of tables. The
usefulness of these tables is that they provide the appropriate Lie
symmetries which can be used for the reduction of the Klein Gordon equation
in Bianchi I\ spacetimes and subsequently the determination of corresponding
invariant solutions.

One important byproduct of this study concerns the wave equation. Indeed the
latter is obtained from the Klein Gordon equation if one considers the
potential to be zero. In this case the constraint condition becomes $\Delta
\psi =0$, that is the conformal factor of the CKVs must be a solution of the
wave equation This means that the point symmetries of wave equation in
Bianchi I spacetimes are the point symmetries of tables \ref{BianchiIG}, \ref%
{BianchiIG4}, \ref{BianchiIGCKV} and \ref{BianchiIGCF}, with potential $%
V\left( t,x,y,z\right) =0$ whereas the vector fields of table \ref%
{BianchiIGCF2} are not Lie /Noether point symmetries of the wave equation
because the conformal factor of the CKVs is not a solution of the wave
equation.

Concerning the reduction procedures which we considered in section \ref%
{KGBianchi} they remain valid for the wave equation. An exact solution of
the wave equation in Kasner universe has been found recently in \cite%
{Esposito}. We would like to note that the ansantz of \cite{Esposito} it is
based on the group invariants of the wave equation (\ref{KGL.01}) as given
in section \ref{casea1}.

A study of the Klein Gordon equation and the symmetries of classical
particles in other Bianchi type spacetimes is under investigation and the
results will be presented in a subsequent paper.

\section*{Acknowledgments}

The authors thank the referee for the useful remarks which improved this
work. AP acknowledge financial support of INFN.


\begin{thebibliography}{99}
\bibitem{ovsiannikov} L. V. Ovsiannikov, \textit{\ Group analysis of
differential equations}, Academic Press, New York, (1982).

\bibitem{Ames81} W. F. Ames, R. J. Lohner and E. Adams, Group properties of $%
u_{tt}=(f(u)u_{x})_{x}$, Int. J. Non-Linear Mech. {16}, 439 (1981).

\bibitem{TsamNLD} M. Tsamparlis and A. Paliathanasis, Lie symmetries of
geodesic equations and projective collineations, Nonlinear Dynamic, 62,
(2010), 203

\bibitem{Camci2014} U. Camci, Symmetries of geodesic motion in G\"{o}%
del-type spacetimes, JCAP 07 (2014) 002

\bibitem{Damianou2d} P.A. Damianou and C. Sophocleous, Symmetries of
Hamiltonian systems with two degrees of freedom, J. Math. Phys. 40 (1999) 210

\bibitem{Damianou3d} P.A. Damianou and C. Sophocleous, Classification of
Noether Symmetries for Lagrangians with Three Degrees of Freedom, Nonlinear
Dynamics 36 (2004) 3

\bibitem{Tsam2d} M. Tsamparlis and A. Paliathanasis, Two-dimensional
dynamical systems which admit Lie and Noether symmetries, J.Phys.A: Math.
Theor, 44, (2011), 175202, (arXiv:1101.5771)

\bibitem{Tsam3d} M. Tsamparlis, A. Paliathanasis and L. Karpathopoulos,
Autonomous three-dimensional Newtonian systems which admit Lie and Noether
point symmetries, J. Phys. A: Math. Theor. 45 (2012) 275201,
(arXiv:1111.0810)

\bibitem{Ibragimov1991} N.H. Ibragimov, M. Torrisi and A. Valenti,
Preliminary group classification of equations $%
v_{tt}=f(x,v_{x})v_{xx}+g(x,v_{x})$, J. Math. Phys., 32, 2988 (1991).

\bibitem{Torrisi1998} M. Torrisi and R. Tracina, Equivalence transformations
and symmetries for a heat conduction model, Internat. J. Non-Linear Mech.
33, 473 (1998).

\bibitem{Zhdanov1999} R.Z. Zhdanov and V.I. Lahno, Group classification of
heat conductivity equations with a nonlinear source, J. Phys. A: Math. Gen.
32, 7405 (1999).

\bibitem{Gandarias2004} M.L. Gandarias, M. Torrisi and A. Valenti, Symmetry
classification and optimal systems of a non-linear wave equation, Int. J.
Non-Linear Mech. 39, 389 (2004).

\bibitem{Ivanova2004} R.O. Popovych, N.M. Ivanova, New results on group
classification of nonlinear diffusion convection equations, J. Phys. A:
Math. Gen. 37, 7547 (2004).

\bibitem{Lahno2006} V. Lahno, R. Zhdanov and O. Magda, Group classification
and exact solutions of nonlinear wave equations, Acta Appl. Math. 91, 253
(2006).

\bibitem{Ivanova2006} N.M. Ivanova and C. Sophocleous, On the group
classification of variable coefficient nonlinear diffusion convection
equations, J. Comput. Appl. Math. 197, 322 (2006).

\bibitem{Ivanova2007} D. Huang and N.M. Ivanova, Group analysis and exact
solutions of a class of variable coefficient nonlinear telegraph equations,
J. Math. Phys., 48, 073507 (2007).

\bibitem{Mahomed2007} F.M. Mahomed. Symmetry group classification of
ordinary differential equations: Survey of some results, Math. Methods Appl.
Sci. 30, 1995 (2007).

\bibitem{Vaneeva2008} O.O. Vaneeva, R.O. Popovych and C. Sophocleous,
Enhanced group analysis and exact solutions of variable coefficient
semilinear diffusion equations with a power source, Acta Appl. Math. 106, 1
(2008).

\bibitem{Ivanova2010} N.M. Ivanova, C. Sophocleous and P.G.L. Leach, Group
classification of a class of equations arising in financial mathematics, J.
Math. Anal. Appl. 372, 273 (2010).

\bibitem{Azad2010} H. Azad and M.T. Mustafa, {Group classification, optimal
system and optimal reductions of a class of Klein Gordon equations}, Commun.
Nonlinear. Sci. Numer. Simul. 15, 1132 (2010).

\bibitem{Tracina2010} R. Tracina, N.M. Ivanova and C. Sophocleous, Group
Classification of Three-Dimensional Variable-Coefficient Burgers Equation,
Waves and Stability in Continuous Media, 1, 224 (2010).

\bibitem{TsamQadir} M. Tsamparlis, A.\ Paliathanasis and A. Qadir, Noether
symmetries and isometries of the minimal surface Lagrangian under constant
volume in a Riemannian space, Int. J. Geom. Methods Mod. Phys. 12 (2005)
155003

\bibitem{Chris} T.\ Christodoulakis, N. Dimakis and P.A. Terzis, Lie point
and variational symmetries in minisuperspace Einstein gravity, J. Phys.\ A:
Math. Theor., 47, 095202 (2014)

\bibitem{Jamal1} S. Jamal, A. H. Kara and R. Narain, Wave equations in
Bianchi Space-times, J. App. Math. 2012, 765361 (2012)

\bibitem{Camci} U. Camci, S. Jamal, A.H. Kara, Invariances and Conservation
Laws Based on Some FRW Universes, Int J. Theor. Phys., vol.53, 1483 (2014)

\bibitem{Jamal2} S.\ Jamal, A. H. Kara and A. H. Bokhari,Symmetries,
conservation laws, reductions, and exact solutions for the Klein-Gordon
equation in de Sitter spacetimes, Can. J. Phys. 90, 667 (2012)

\bibitem{Azad2013} H. Azad, Ahmad Y. Al-Dweik, R. Ghanam and M. T. Mustafa,
Symmetry analysis of wave equation on static spherically symmetric
spacetimes with higher symmetries, J. Math. Phys., 54, 063509 (2013).

\bibitem{AnJGP} A. Paliathanasis and M. Tsamparlis, Lie point symmetries of
a general class of PDEs: The heat equation, J. Geom. Phys., 62 (2012) 2443

\bibitem{AnIJGMMP} A. Paliathanasis and M. Tsamparlis, The geometric origin
of Lie point symmetries of the Schr\"{o}dinger and the Klein--Gordon
equations, Int. J. Geom. Methods Mod. Phys. 11 (2014) 1450037

\bibitem{Ibrag} N.H. Ibragimov, \textit{Transformation Groups Applied to
Mathematical Physics}, D. Reidel Publishing Co, Dordrecht (1985).

\bibitem{Stephani} H. Stephani, \textquotedblleft Differential Equations:
Their Solutions Using Symmetry", Cambridge University Press, New York,
(1989).

\bibitem{Govinger} K.S. Govinger, Lie subalgebras, reduction of order, and
group-invariant solutions, J. Math. Anal. Appl. 258 (2001) 720.

\bibitem{Bluman} G.W. Bluman and S. Kumei, Symmetries of Differential
Equations, (Springer-Verlag, New York, (1989))

\bibitem{Katzin} G.H. Katzin, J. Levine and R.W. Davis, Curvature
Collineations: A Fundamental Symmetry Property of the Space-Times of General
Relativity Defined by the Vanishing Lie Derivative of the Riemann Curvature
Tensor, J. Math. Phys. 10, 617 (1969)

\bibitem{Bozhkov} Y. Bozhkov and I.L. Freire, Special conformal groups of a
Riemannian manifold and Lie point symmetries of the nonlinear Poisson
equation, J. Differential Equations 249 (2010) 872

\bibitem{BRyan} M.P.Jr. Rayan and L.C. Shepley, Homogeneous Relativistic
Cosmologies, Princeton University Press, Princeton (1975)

\bibitem{TsAp} M. Tsamparlis and P.S. Apostolopoulos, Symmetries of Bianchi
I space-times, J. Math. Phys., 41 (2000) 7573

\bibitem{GovinderLeach} K.S. Govinder and P.G.L. Leach, An elementary
demostration of the existence of sl(3,R) symmetry for all second-order
linear ordinary differential equations, SIAM Review 40 (1998) 945

\bibitem{Esposito} E. Battista, E. Di Grezia and G. Esposito, Scalar wave
equation in Kasner spacetime, arXiv:1410.3971
\end{thebibliography}
\end{document}